\documentclass[conference, a4paper]{IEEEtran}
\IEEEoverridecommandlockouts
\usepackage{cite}
\usepackage{amsthm,amsmath,amssymb,amsfonts}
\usepackage{algorithmic}
\usepackage{graphicx}
\usepackage{textcomp}
\usepackage{xcolor}
\usepackage{float}
\usepackage{algorithm}
\usepackage{algorithmic}
\usepackage{bbold}
\usepackage{array}
\usepackage{booktabs}
\usepackage{mathrsfs}
\usepackage[normalem]{ulem}

\usepackage{todonotes}
\usepackage{multirow}
\usepackage{makecell}
\usepackage{subfig}

\usepackage[hyphens]{url}
\usepackage{hyperref}
\usepackage[hyphenbreaks]{breakurl}

\def\BibTeX{{\rm B\kern-.05em{\sc i\kern-.025em b}\kern-.08em
    T\kern-.1667em\lower.7ex\hbox{E}\kern-.125emX}}

\IEEEaftertitletext{\vspace{-1.16\baselineskip}}
\begin{document}

\title{Adaptive Dynamic Programming for Energy-Efficient Base Station Cell Switching\\
{}
}

\author{
    \IEEEauthorblockN{Junliang Luo, Yi Tian Xu, Di Wu, Michael Jenkin, Xue Liu, Gregory Dudek}
    \IEEEauthorblockA{Samsung AI Center Montreal, Canada
    \\\{firstname.lastname\}@samsung.com, di.wu1@samsung.com}
    }


\maketitle

\begin{abstract}
Energy saving in wireless networks is growing in importance due to increasing demand for evolving new-gen cellular networks, environmental and regulatory concerns, and potential energy crises arising from geopolitical tensions.
In this work, we propose an approximate dynamic programming (ADP)-based method coupled with online optimization to switch on/off the cells of base stations to reduce network power consumption while maintaining adequate Quality of Service (QoS) metrics. 
We use a multilayer perceptron (MLP) given each state-action pair to predict the power consumption to approximate the value function in ADP for selecting the action with optimal expected power saved.
To save the largest possible power consumption without deteriorating QoS, we include another MLP to predict QoS and a long short-term memory (LSTM) for predicting handovers, incorporated into an online optimization algorithm producing an adaptive QoS threshold for filtering cell switching actions based on the overall QoS history.
The performance of the method is evaluated using a practical network simulator with various real-world scenarios with dynamic traffic patterns.
\end{abstract}

\begin{IEEEkeywords}
    Energy Saving, Base station sleeping, Switching cell on/off, Approximate Dynamic Programming.
\end{IEEEkeywords}

\section{Introduction}
Energy consumption associated with global 5G infrastructure has reached an unprecedented scale \cite{reports_on_increased_energy_consumption}.
%
Compared to their 4G counterparts, base stations in 5G networks require massive amounts of energy to operate since 5G base stations must be more densely deployed than 4G base stations due to the fact that 
the high-frequency bands associated with 5G networks have reduced converge compared to the lower-frequency bands associated with 4G\cite{wisely2018capacity}.
Improving energy efficiency can help  a reduction in global greenhouse gas emissions \cite{freitag2021real,wu2017two}, cost saving for the telecommunication operators \cite{srivastava2020energy}, compliance with the energy regulations in the telecommunication industry \cite{saelens2019impact}, and reduced risk associated with global energy instability \cite{delardas2022ripple}.
%
%

%
Algorithms for limiting energy consumption in wireless networks
have received significant attention from the research community leading to some promising results \cite{salahdine2021survey}.
In earlier works,  energy saving is typically formulated as an optimization problem with the goal of optimizing the energy efficiency of the data transmission process \cite{li2014energy}, optimizing the positioning of base stations during  deployment \cite{coskun2014energy}, and optimizing the on/off status of base stations \cite{feng2017base}, etc.
Base station switching, i.e., selectively switching off components with low usage in base stations, is considered to be a flexible solution due to its fast adaptability without requiring to change any physical components in base stations \cite{wu2015energy}, and the fact that many geographical locations are served by one or more frequency bands delivered by one or more base stations. Turning off entire base stations or certain frequency bands is 
feasible and applicable since base stations are typically deployed and configured to satisfy peak traffic loads, and in normal operation the maximum usage is not always reached \cite{feng2017base}.
The challenge is to find a strategy to maximize the energy saved without deteriorating the quality of service (QoS) delivered to the user.
Switching off base station components can impact the network performance such as increasing the transmission delay and creating network congestion. To address the problems, some works consider degrading communication delay \cite{son2011base} or the traffic load power consumption ratio \cite{tan2022graph}. 
%
%
Network performance can be considered as a continuous measure, or as a constraint that the performance not be permitted to degrade outside of a predefined acceptable range. 
A range of studies exist in the literature.
Niu et al. \cite{niu2015characterizing} proposed a sleep mode strategy named N-Policy to set the base stations to sleep mode when utilization is low over a time period.
Salem et al. \cite{salem2017advanced} performed base station switching between active and sleep modes periodically and maximized the sleep mode duration according to the maximum communication delay tolerance to reduce power consumption.
Tan et al. \cite{tan2022graph} proposed a GNN method to model the network topology into a graph where each base station serves as a node with traffic conditions as the node features. An embedding vector is learnt for each node, which is subsequently mapped to an on/off switching decision.
Pujol–Roigl et al. \cite{pujol2021deep} proposed a deep reinforcement learning (RL) method combining energy consumption, IP throughput rate, and handover in the reward function.
While these and related works offer various strategies for base station switching, they either reply on pre-defined performance degradation constraints that potentially overlook the real-time change in traffic \cite{niu2015characterizing, salem2017advanced}, or involve a complex reward or loss function that is computationally heavy \cite{tan2022graph, pujol2021deep}.
We tackle those problems using an ADP algorithm utilizing prior knowledge about the environment, including traffic and QoS patterns, to have adaptability and computational efficiency.
%
%
%
ADP algorithm can be more computationally efficient and more sample efficient for small action spaces, compared to RL methods and faster training and convergence for the value function approximation \cite{xu2018learning}.
We formulate the energy saving problem as a sequence of decision-making processes involving the selection of cells to switch off within a dynamic traffic environment, while considering the trade-off between the conflicting objectives of minimizing power consumption and maintaining the QoS.
Experiments conducted on  on a practical network simulator show that the proposed ADP method can achieve up to a 20.1\% of power saving compared to baseline energy consumption.
Compared to existing approaches, the main advantages of the proposed ADP-based energy saving method presented in this paper are:
\begin{enumerate}
    \item \textit{Neural Networks-based estimators} The proposed ADP method employs two multilayer perceptron (MLP) estimators and a LSTM \cite{hochreiter1997long} estimator, trained on widely distributed historical traffic and action data, to perform power consumption, QoS and handovers estimation, for an proper approximation of the value function.
    
    \item \textit{Adaptive Target Online Adaptation:} An online optimization process for QoS constraint adjustment is employed. Utilizing real-time cumulative QoS records during interactions with the environment to determine an adaptive target QoS, allows for responsiveness to fluctuating network conditions for optimizing energy savings while maintaining desired QoS levels in a wireless network.

    \item \textit{Broad Applicability:} In addition to the simulated environment presented in this paper, the method can be applied to various wireless network configurations, accommodating different traffic scenarios and QoS metrics following the identical pipeline.
\end{enumerate}

\section{System Description} %
\begin{figure}[!t]
    \centering
    \includegraphics[width=0.83\linewidth]{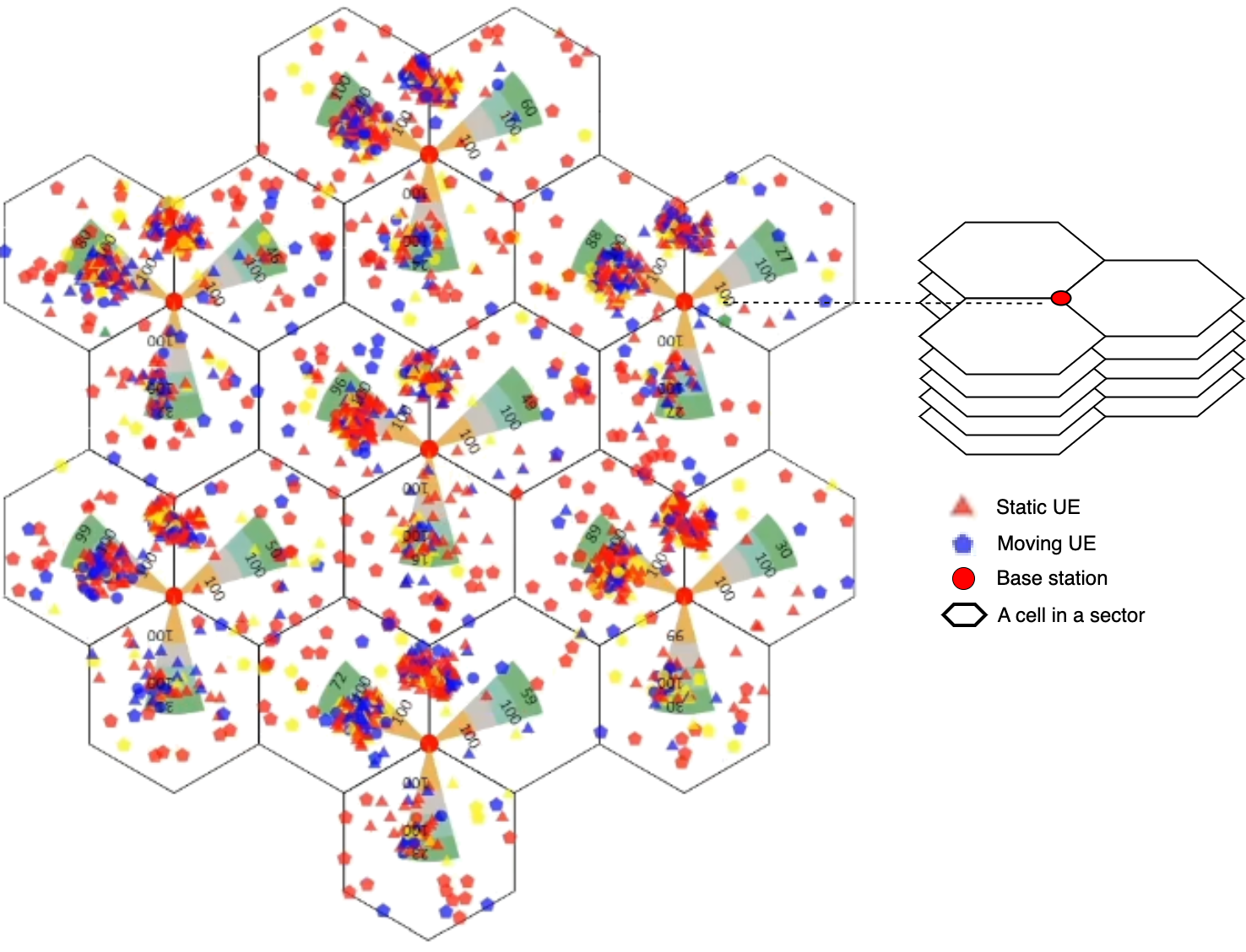}
    \caption{Spatial layout of a base station with three sectors and five cells per sector, showcasing the simulated network topology and UEs of the simulator.}
    \label{fig_sls}
\end{figure}
We consider a wireless network that consists of multiple base stations $B_{1},...,B_{K}$ distributed over an area in a hexagonal layout as shown in Figure. \ref{fig_sls}. 
Each base station serves UEs in three sectors (hexagonal regions) and each sector has five frequency carriers each of which corresponds to a cell. The same frequency carriers are shared for the sectors of all the base stations.
We denote a cell, i.e., a particular carrier in a particular base station as $C_{k,i,j}$: the cell $j$ at the sector $i$ of the base station $B_k$.
The load ratio of a cell at the time step $t$ is denoted as $\lambda^{t}_{k, i, j} = \frac{N^{t}_{k, i, j}}{M_j} \in [0,1]$, where $M_j$ is the maximum total physical resources blocks (PRBs) of a cell $j$, which is the same for all cells of the same frequency. $N^{t}_{k, i, j}$ is the total PRBs allocated to the users at time $t$ of the cell $C_{k,i,j}$.
The power consumption of a cell $C_{k,i,j}$ given a load ratio $\lambda^{t}_{k, i, j}$ is expressed as:
\begin{equation}
    P^{t}_{k, i, j} = e^{t}_{k, i, j} \cdot (P1_{j} + \lambda^{t}_{k, i, j} P2_{j})  + (1 - e^{t}_{k, i, j}) \cdot P0_{j} + \Delta_{t},
\end{equation}
where $e^{t}_{k, i, j} = 1 $ when $C_{k,i,j}$ is on and $e^{t}_{k, i, j} = 0$ when $C_{k,i,j}$ is off at the time step $t$.
In our system, $e^{t}_{k, i, 0}$ is set to be $1$ at all times, which corresponds to the cell with the lowest frequency, therefore, the largest coverage range. This constraint is set to guarantee service coverage.
$P1_{j}$ represents the standby power consumption of an active cell $j$ even with zero traffic load, $P2_{j}$ represents the power consumption that scales linearly with the current load ratio of the cell, and $P0_{j}$ is sleeping power consumption when a cell is off.
$\Delta_{t}$ is the power consumption associated with switching on a cell. 
$\Delta_{t} = \beta \cdot P_{\gamma} \mathbb{1}$, where $\beta, P_{\gamma}$ are constants, $\mathbb{1}(x = e^{t}_{k, i, j} - e^{t-1}_{k, i, j}) = 1 \enspace \text{if} \enspace x=1 \enspace \text{else} \enspace 0$.

The on/off switching action $u^{t}_{k}$  changes the value of $e^{t}_{k, i, j}$  given the current state $X^{t}_{k}$.
%
%
The current on/off status is represented by $e^{t}_{k} = [e^{t}_{k, 0, 0}, ..., e^{t}_{k, 2, 4}]$, i.e., the status for each cell of the base stations $k$ made by last action $u^{t-1}_{k}$.
%
%
The state $X^{t}_{k}$ of a base station $k$ at the time step $t$ is defined as:
$X^{t}_{k} = [UE^{t}_{k}, \ TP^{t}_{k}, \ N^{t}_{k}, \ e^{t}_{k}]$,
a tuple of active UEs, IP throughput, cell PRB load ratio, and current on/off status. 
We represent the QoS ($Q^{t}_{k}$) in terms of the percentage of uncongested cells. 
\begin{equation}
    Q^{t}_{k} = \frac{\sum_{i, j} \{ c = 1 \ \text{if} \ \frac{D^{t}_{i, j}}{{TT}^{t}_{i, j}} < \tau \ \text{else} \ 0 \}}{\sum_{i, j} \{ c=1 \} \enspace \text{if} \enspace \scalebox{0.8}{$TT_{i,j}$} > 0} 
\end{equation}
The percentage of uncongested cells  at time step $t$ is defined as the percentage of the active cells with the amount of successfully transmitted data ($D^{t}_{i, j}$) over the transmission time ($TT^{t}_{i, j}$), lower than a threshold $\tau$ (e.g., 1 Mbps).
\begin{equation}
    H^{t}_{k} = \sum^{3}_{i=1}\sum^{5}_{j=1} \frac{|UE^{t-1}_{k,i,j} - UE^{t}_{k,i,j }|}{2}
\end{equation}
The handover of a base station $k$ at the time step $t$ is definedas in \cite{pujol2021deep}, where $UE^{t}_{k,i,j}$ stands for the user equipment ($UE$) connected to $C_{k,i,j}$ at the time step $t$.
\section{Problem Formulation and Methods}
The energy saving problem of base station cell switching is formulated as a Markov decision process (MDP).
%
The formulation is chosen since the problem involves sequential decision-making. 
The decision of switching happens at every discrete time step in a stochastic environment, where the future UE movement and demands are unknown and a trade-off is needed between minimizing the energy consumption and maintaining acceptable QoS.
%
%
In RL solutions such as \cite{salem2018reinforcement, pujol2021deep}, the trade-off is considered by a reward function that combines both the power consumption and QoS quality such as the IP throughput in \cite{pujol2021deep}.
In the ADP-based solution considered, this trade-off is addressed through 
a constraint-based approach that treats the QoS requirement as a constraint that enforces the constraint satisfaction \cite{chakrabarty2019approximate}.
We adopt the problem formulation given in \cite{chakrabarty2019approximate} with the modification of switching on/off at the cell-level.
In this MDP formulation, our algorithm is applied on each base station individually (e.g., for a base station $k$).
The optimal policy, $\pi^*$ minimizes the expected cumulative power consumption over $T$ time steps and satisfies the QoS constraint.
\begin{equation}
    \begin{aligned}
        \pi^* = \arg\min_\pi \mathbb{E}\left[
        \sum_{t=0}^{T} C(X^{t}_{k}, \pi(X^{t}_{k}))
        \right] 
        \text{s.t.} \\
        \forall t, \enspace \mathbb{E}\left[ Q(X^{t}_{k}, \pi(X^{t}_{k})) \right] > Q_{\tau}
    \end{aligned}
    \label{cost_to_go}
\end{equation}
Under a dynamic programming setting, we have a cost-to-go value function denoted as $J_t(X_{k}^{t})$ representing the expected minimal total cost of completing this energy saving problem from a given time step $t$ until the last time step T to solve to obtain the optimal policy iteratively. 
\begin{equation}
    \begin{aligned}
        J_t(X_{k}^{t}) = \min_{u^{t}_{k}} \mathbb{E} \left[
        C(X^{t}_{k}, u^{t}_{k}) + J_{t+1}(X^{t+1}_{k}, u^{t}_{k}) 
        \right] 
        \text{s.t.} \\
        \forall t, \enspace \mathbb{E}\left[ Q(X^{t}_{k}, u^{t}_{k})) \right] > Q_{\tau}
    \end{aligned}
    \label{cost_to_go}
\end{equation}
Assuming the traffic load occupies a finite space denoted by $\Lambda$, which is anticipated to have a considerable size. The cost-to-go function (equation (\ref{cost_to_go}), presents a state and action space of $ T \times |\Lambda| \times |U| $. $|U|$ represents the action set, highlighting the array of feasible actions within the system. Exhaustive search of this configuration leads to computational challenges.
In approximate dynamic programming, the cost-to-go value function is approximated by less complex and  more computationally efficient functions to deal with the large state and action space. 
The approximation allows to find near-optimal solutions to the problem and the methods of approaching include kernel-based function and neural networks-based functions \cite{deptula2018approximate}.

\subsection{ADP Neural network-based Estimators}
Given the traffic load $X^{t}_{k}$ and possible on/off switching action $u^{t}_{k}$ at time $t$, we use three neural network-based estimators for approximation:
\textit{Power Consumption Estimator}: 
$\Tilde{P}(X^{t}_{k}, u^t_k)$; 
\textit{QoS Estimator}: 
$\Tilde{Q}(X^{t}_{k}, u^t_k)$; 
\textit{Handover Prediction Estimator}:
$\Tilde{H}(X^{t}_{k}, e^{t-1}_{k}, u^t_k)$.
A multi-layer perceptron  (MLP) is employed to predict the power consumption, while a second MLP predicts the QoS represented by the ratio of uncongested cells given all possible pairs of $[X^{t}_{k}, u^{t}_{k}], \forall u^{t}_{k} \in U$ at each time step $t$. A long short-term memory (LSTM) model  is used to predict the handover with the additional the cells on/off state of the last time step ($e^{t-1}_{k}$) as the input. 
%
%
%
The estimators were trained with trained with supervised learning on the historical data containing a sufficiently rich combination of the input states and actions, and the output ground truth of power consumption, QoS, and handovers.
The records of power consumption, QoS, and handover records the estimators trained on,  were obtained from data generated across 64 runs for each of the 8 scenarios. In every individual run, random actions were executed throughout the 96 time steps.
The optimal action $u^{t*}_{k}$ is selected as the action with the lowest predicted power consumption among all the actions with predicted QoS above the QoS constraint $Q_{\tau}^t$.
\begin{equation}
    u^{t*}_{k} = \arg\min_{\mathbf{u}^t_k} (\Tilde{P}^t_k) \quad \text{s.t.} \quad \Tilde{Q}^t_k \ge Q_{\tau}^t
\end{equation}
The dynamic QoS threshold, $Q_{\tau}^t$ is derived from an adaptive function that integrates predicted handovers and current QoS for online optimization. 
Utilizing a static threshold can be problematic due to the risk of constant oscillations, particularly when actions like switching a cell off and on repeatedly breach and then fall below the threshold (ping pong effect). 
By factoring in predicted handovers, the dynamic threshold enables stricter QoS constraints when a cell deactivation is predicted to trigger significant handovers given the current traffic loads. 
When a cell deactivation is predicted to induce significant handovers, the threshold becomes stricter, thereby avoiding the repetitive toggling of cells on and off. 
Also, during low-traffic periods, when more actions of switch offing additional cells are predicted with to have low predicted handovers, the dynamic threshold is adjusted to facilitate greater energy efficiency.
%
%

\subsection{Adaptive Target Online Optimization}
Given the average (on all the actions) predicted handover $\Bar{H}^t$, the parameters $\theta^*$ are updated to minimize the difference between the target QoS $Q^t_{\phi}$, and the threshold $Q_{\tau}^{t}$.
\begin{equation} 
    \theta^* = \arg\min_\theta [(Q^t_{\phi} - Q_{\tau}^{t})^2 + \gamma ||\theta||^2 ],
\end{equation} 
where $\theta$ represents the adjustable parameters of a function $g_\theta$ models the relationship between the predicted average handovers $\Bar{H}^t$ and the adaptive QoS threshold $Q_{\tau}^{t}$.
$\gamma ||\theta||^2$ is a regularization squared Euclidean norm of the parameter vector to prevent overfitting.
\begin{equation} 
Q_{\tau}^{t} = g_{\theta}(\Bar{H}^t) = \theta_0 + \theta_1 \Bar{H}^t
\end{equation} 

We use a linear function $g_\theta$ with an objective function updated by gradient descent. 
$\theta_0$ represents the intercept (the QoS threshold when no handovers), and $\theta_1$ represents the slope (how the QoS threshold changes with respect to the handover). The bounds for $\theta_0, \theta_1$ for the optimization to yield practical results are presented in Table. \ref{table:parameters}.
\begin{equation}
L(\theta) = (Q^t_{\phi} - (\theta_0 + \theta_1 \Bar{H}^t))^2 + \gamma ||\theta||^2
\end{equation}
The adaptive target QoS $Q^t_{\phi}$ is calculated as: $Q^t_{\phi} = Q_{\Phi} + Q^t_{\delta}$, where $Q_{\Phi}$ denotes a constant target QoS, and $Q^t_{\delta}$ represents the difference between the target QoS and the average of observed QoSs up to the current $t$:
$Q^t_{\delta} = Q_{\Phi} - \frac{1}{t} \sum_{i=1}^{t} Q^i$.
$Q^t_{\delta}$ is positive/negative when the observed QoS is lower/higher than the target QoS, which will adjust  $Q^t_{\phi}$ accordingly. 
Essentially, $\theta_0$ and $\theta_1$ learned at time step $t$ are utilized along with the handovers at the next step for determining the action. 
After each action, parameters $\theta_0$ and $\theta_1$ are recalibrated based on \(Q^t_{\phi}\). This adjustment, informed by all observed QoS records, continues to balance energy savings with QoS upkeep during online optimization.

\subsection{Certainty Equivalent Control and Cost-to-Go Estimation}
In addition to the aforementioned estimators that are trained on historical data to estimate the power consumption and QoS, we apply the Certainty Equivalent Control (CEC) in \cite{ayala2018energy, bertsekas1996dynamic}, which replaces the $J_{t+1}(X^{t+1}_{k}, u^{t}_{k})$ in Equation. \ref{cost_to_go} with its expected value $\Tilde{J}_{t+1}(\Bar{X}^{t+1}_{k}, u^{t}_{k})$ to further simplify the control strategy.
CEC transfers the stochastic state transition in the next time step cost $J_{t+1}$ to be deterministic to further reduce the computational cost.
$\Bar{X}^{t+1}_{k}$ is the mean traffic from historical data at a certain time step using the estimators to store a table of $\Tilde{J}_{t}(\Bar{X}^{t}_{k}, u^{t}_{k}), \forall t,  u^{t}_{k} \in U$, i.e., an offline phase of ADP.
The mean traffic of historical data is used due to a reasonable assumption that the user traffic demands follow certain periodic patterns throughout a day.
\addtolength{\topmargin}{0.08in}
In the offline phase, the expected cost-to-go function is calculated as the following, where $Q_{\tau \prime}$ is a constant, $\Delta_{u^{t}_{k}}$ is the on/off state transition cost associated with switching on additional cells.
\begin{equation}
    \begin{aligned}
        \Tilde{J}_{t}(\Bar{X}^{t}_{k}, u^{t}_{k}) = 
        \min_{u^{t}_{k}} [ \Tilde{P} (\Bar{X}^{t}_{k}, u^{t}_{k}) +
        \Tilde{J}_{t+1}(\Bar{X}^{t+1}_{k}, u^{t}_{k})  \\ +
        \Delta_{u^{t}_{k}} ] \enspace \text{s.t.} 
        \quad \Tilde{Q}^t_k \ge Q_{\tau \prime}  
    \end{aligned}
    \label{Eq:cost-to-go_functiohn}
\end{equation}
Upon completing the offline training, the online phase can commence.
The ADP algorithm for energy-efficient base station switching can be performed on scenarios of on-going traffics, while continuously adapting to the changing network traffic conditions.
The online phase is stated as:
\begin{equation}
    \begin{aligned}
    u^{t*}_{k} = \arg\min_{u^{t}_{k}} [ \Tilde{P} (X^{t}_{k}, u^{t}_{k}) +
        \Tilde{J}_{t+1}(\Bar{X}^{t+1}_{k}, u^{t}_{k}) \\ + 
        \Delta_{u^{t}_{k}} ] \enspace \text{s.t.} 
        \enspace \Tilde{Q}^t_k \ge Q_{\tau}^{t}=g_{\theta^*}(\tilde{H}^{t}_{k})
    \end{aligned}
    \label{Eq:optimal_action}
\end{equation}
\section{Experiments}\label{exp}

Experiments were performed using a 5G proprietary system-level network simulator described in \cite{li2022traffic}. The simulator was configured using  preset traffic scenarios obtained from real-world traffic historical data.
%
%
UEs are simulated as points that are either concentrated uniformly at specified regions in a sector or uniformly distributed across the entire environment.
The simulated network is as shown in Figure~\ref{fig_sls}. Each base station has three sectors each covering a hexagonal geographical area with five cells (carrier bands).
In Figure~\ref{fig_sls} the colour of each UE represents its connected band (cell).

\begin{table}[!htbp] 
\centering
\caption{Parameters of Simulator}
\resizebox{0.376\textwidth}{!}{
\begin{tabular}{ll}
\hline
\textbf{Parameter} & \textbf{Values}\\
\hline
Number of scenarios & 8 \\
Running time of the simulation & 1 day \\
Number of steps & 96 (15-min a step) \\
hexagonal geographical areas & Hex 1 Individual \\ 
Number of sectors per base station & 3 \\
Number of cells per sector & 5 (1. always on) \\
\hline
\end{tabular}
}
\label{table_simulator_parameters} 
\end{table}

The parameters used in the simulator are reported in Table~\ref{table_simulator_parameters}.
We employ a one-day simulation runtime to evaluate the performance of the proposed algorithm as this duration yields  experimental results for the representative daily test scenarios.
A single experiment duration consists of 96 discrete time steps each corresponding to 15 minutes.
The current traffic state including active UEs, cell PRB, IP throughput, and cell on/off status of all the cells are input to the ADP algorithm at each step for deciding an action for the net time tep. 
The action $u^{t}_{k}$ taken at the time step $t$, along with the traffic of active UEs of the scenarios in the simulation, influences the next time step state $X^{t+1}_{k}$.
\begin{table}[!htbp]
\centering
\caption{Parameters in System}
\resizebox{0.497\textwidth}{!}{
\begin{tabular}{lll}
\hline
\textbf{Parameter} & \textbf{Description} & \textbf{Value} \\
\hline
$\beta, P_{\gamma}$ & \makecell[l]{Power consumption for switching on $n$ \\  inactive base stations: $\Delta = \beta \cdot P_{\gamma} \cdot n$}  & 0.3, 162\\
$\tau$ & Threshold of the ratio of uncongested cells calculation  & 1 Mbps \\
$Q_{\tau, \prime}$ & QoS constant threshold for offline cost-to-go table generation & 80 \\
$Q_{\Phi}$ & Constant for the target QoS & 92 \\
$m$ & Number of runs for generating random actions for training the estimators & 64 \\
$\theta_0, \theta_1$ & Bounds for the optimization factors & [80,90], [0,3] \\
$\gamma$ & Factor of the regularization norm in online optimization  & 0.001 \\
\hline
\end{tabular}}
\label{table:parameters}
\end{table}
The ADP algorithm is applied to each individual base station. 
%
Total power consumption is the sum of the power of the individual base stations.
Our forthcoming experimental results pertain specifically to one single base station to distill the effect of the proposed ADP algorithm.
The parameters used in the ADP algorithm are reported in Table~\ref{table:parameters}.
The hyperparameters used in the training of the MLP models for power consumption and QoS prediction and the LSTM model for handover prediction are detailed in Table \ref{tab:model_hyperparameters}.

\begin{table}[!htbp]
\centering
\caption{Hyperparameters of Estimators}
\label{tab:model_hyperparameters}
\resizebox{0.397\textwidth}{!}{
\begin{tabular}{lll}
\hline
\textbf{Parameter}   & \textbf{MLP Model} & \textbf{LSTM Model} \\ \hline
Hidden Layers        & 5                  & 3                    \\
Units per Layer      & 64, 128, 128, 64   & 64, 32, 32      \\
Activation Functions & ReLU                & built-in tanh        \\
Optimizer            & Adagrad            & Adagrad              \\
Learning Rate        & 0.001       & 0.05          \\
Loss Function        & MSE                & MSE                  \\ \hline
\end{tabular} 
}
\end{table}

\begin{figure*}[!htbp]
\begin{center}
    \subfloat[]{\includegraphics[width = 0.331\linewidth]{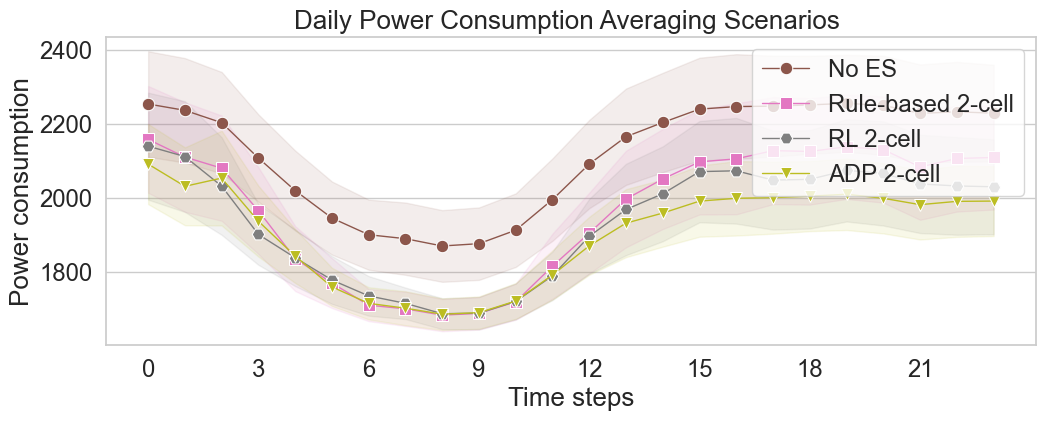}}   
    \subfloat[]{\includegraphics[width = 0.331\linewidth]{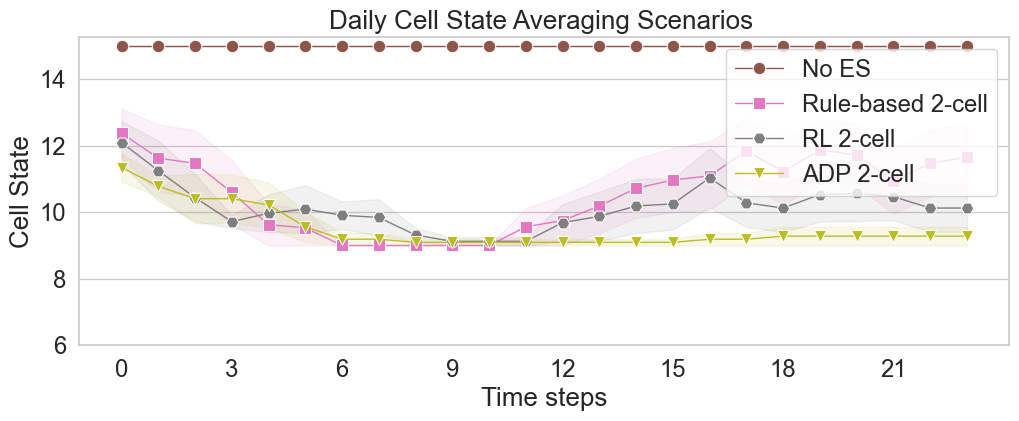}}  
    \subfloat[]{\includegraphics[width = 0.331\linewidth]{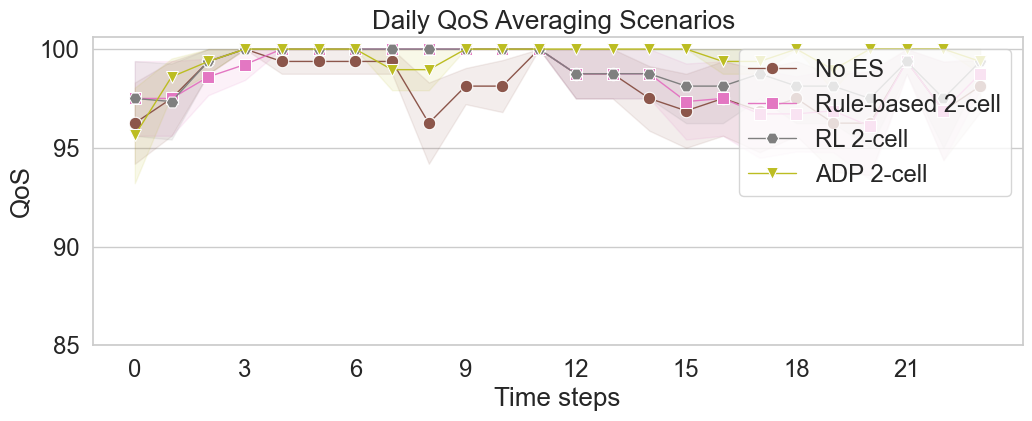}}  \\ \vspace{-0.43cm}
    \subfloat[]{\includegraphics[width = 0.331\linewidth]{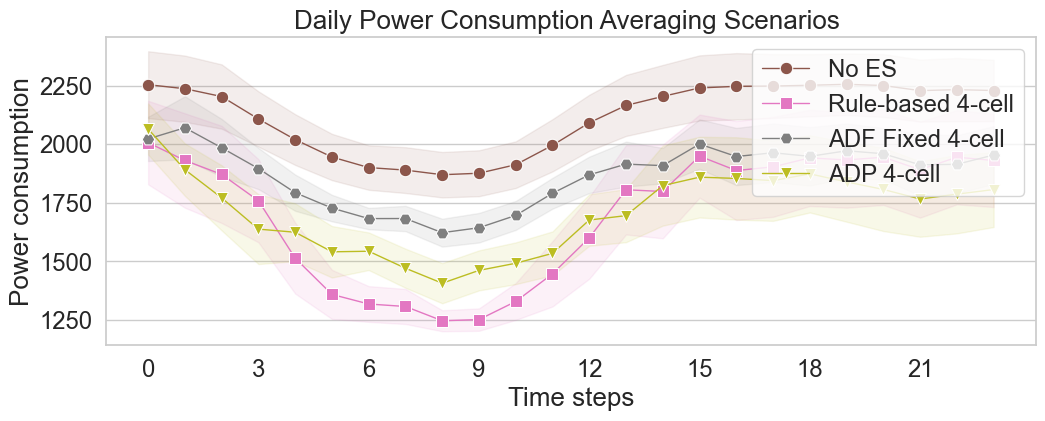}}  
    \subfloat[]{\includegraphics[width = 0.331\linewidth]{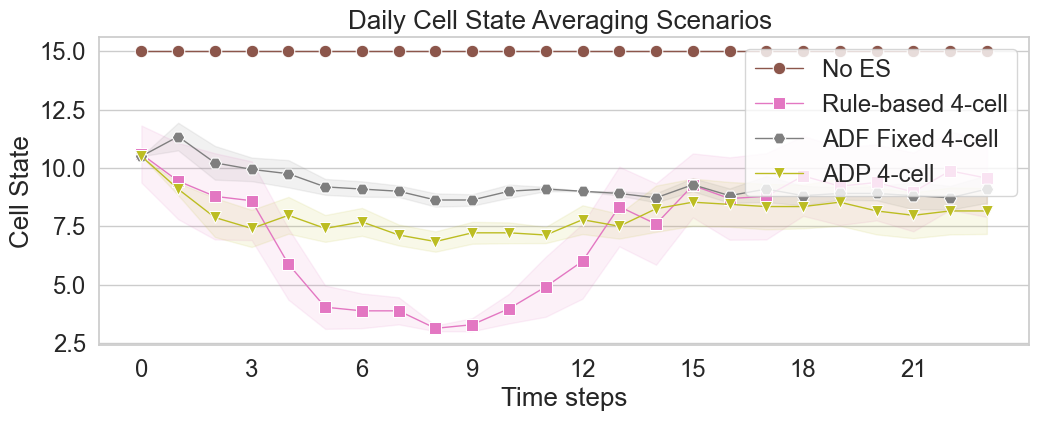}}  
    \subfloat[]{\includegraphics[width = 0.331\linewidth]{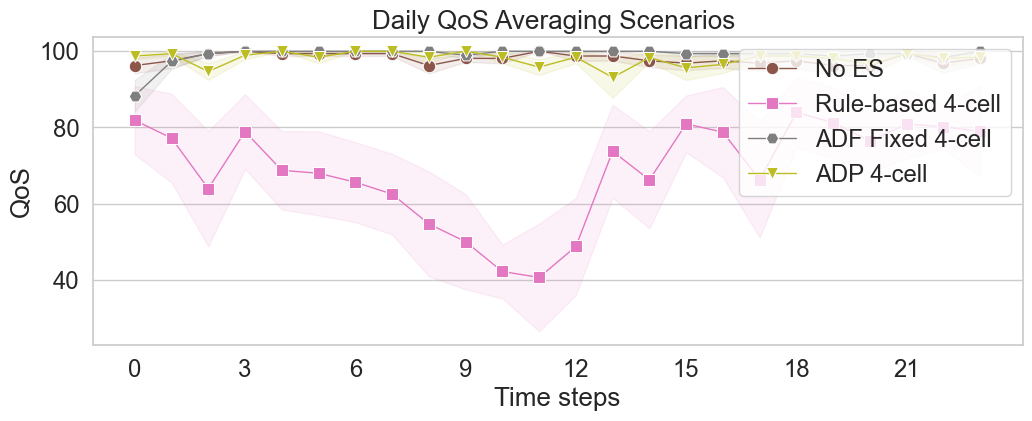}} \vspace{-0.23cm}
\end{center}
\caption{Comparison of power consumption, cell states, and QoS results. Lines and shaded regions denote hourly (aggregate every four 15-minute time steps to reduce visual clutter for clarity) means and standard errors across all scenarios.}
\label{fig:methods_comparison}
\end{figure*}

\begin{figure}[!htbp]
    \centering
    \includegraphics[width=0.97\linewidth]{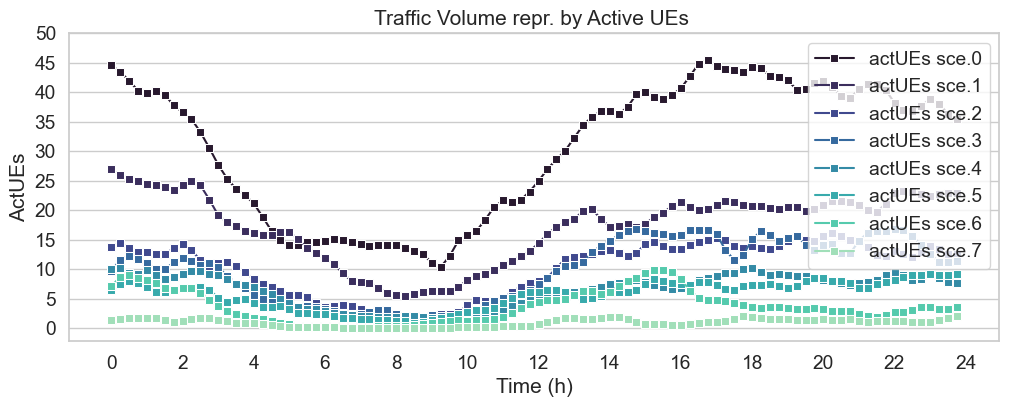}
    \caption{Total active UEs over all cells in eight testing scenarios: The scenarios are ordered by traffic volume magnitude.} 
    \label{fig_traffic}
\end{figure}

As shown in Figure~\ref{fig_traffic}, the traffic demands in the eight scenarios, though distinct in total traffic loads,  follow a similar periodic patterns that consist of peaks and troughs for daily 5G communication representing real-world scenarios.
To gather the training data for our estimators, we load each scenario in the simulator, selecting a random action at every time step. This procedure is repeated for the same scenario  $m$ (64) times to collect samples.
The data generation approach aims at enabling the acquisition of sufficient dynamic actions from data for the purpose of training the estimators without brute forcing all the possible actions.
The probability for one action was never selected is in a sector $(1 - \frac{1}{2^{5-1}})^{64} \approx 1.6\%$ (4 cells in one sector as the first cell is always on). In the simulation, the same action will be applied to all three sections.
To assess our ADP algorithm, we apply the algorithm to all eight scenarios using the simulator and compare the proposed algorithm with three other algorithms:
\begin{itemize} 
\item \textit{No Energy Saving (No ES)}: No energy-saving techniques will be applied. All cells will stay active during the duration of all the steps.
\item \textit{Rule-based cells off (Rule-based)}: Switch the cells on/off using predefined rules based on two thresholds: the deactivation threshold ($th_{deac}$ = 0.2) determines the average load below which a cell should be turned off, and the activation threshold ($th_{ac}$ = 0.8) determines in turning on the inter-sector coexisting cells that are off when the traffic load active cells beyond this threshold.
\item \textit{RL-based method (RL)}: A reinforcement learning method \cite{schulman2017proximal}: Proximal Policy Optimization (PPO) with rewards based on the minimum IP throughput and normalized power consumption equally weighted. 
\end{itemize}

\begin{table}[!htbp]
\centering
\addtolength{\tabcolsep}{0.6pt}
\caption{Performance Results Comparison}
\label{tab:experiment_results}
\resizebox{0.4999\textwidth}{!}{ 
\begin{tabular}{llllllllllll}
\hline
\multicolumn{1}{l|}{Algorithm} & \multicolumn{1}{l|}{Metric} &  & \multicolumn{8}{c}{Scenario} &  \\ \cline{3-12} 
\multicolumn{1}{l|}{} & \multicolumn{1}{l|}{} &  & \multicolumn{1}{c}{1} & \multicolumn{1}{c}{2} & \multicolumn{1}{c}{3} & \multicolumn{1}{c}{4} & \multicolumn{1}{c}{5} & \multicolumn{1}{c}{6} & \multicolumn{1}{c}{7} & \multicolumn{1}{c}{8} & Avg. \\ \hline
\multicolumn{1}{l|}{\multirow{3}{*}{No ES}} & \multicolumn{1}{l|}{Power} &  & 1862 & 2523 & 1738 & 2051 & 1947 & 2113 & 2316 & 2581 & 2141 \\
\multicolumn{1}{l|}{} & \multicolumn{1}{l|}{QoS} &  & 98.9 & 99.5 & 99.7 & 99.3 & 99.5 & 95.2 & 99.7 & 92.1 & 98.0 \\
\multicolumn{1}{l|}{} & \multicolumn{1}{l|}{Handover} &  & 19.0 & 43.7 & 49.4 & 62.7 & 68.4 & 88.3 & 129.2 & 227.0 & 85.9 \\ \hline
\addlinespace[-1.6pt]
 &  &  &  &  &  &  &  &  &  &  &  \\ \hline
\multicolumn{1}{l|}{\multirow{3}{*}{Rule-based 2-cell}} & \multicolumn{1}{l|}{Power} &  & 1565 & 1950 & 1677 & 1915 & 1861 & 2006 & 2233 & 2529 & 1967 \\
\multicolumn{1}{l|}{} & \multicolumn{1}{l|}{QoS} &  & 100.0 & 100.0 & 100.0 & 100.0 & 99.7 & 96.6 & 99.7 & 92.2 & 98.5 \\
\multicolumn{1}{l|}{} & \multicolumn{1}{l|}{Handover} &  & 25.6 & 71.2 & 103.5 & 100.7 & 155.8 & 177.6 & 172.9 & 249.8 & 132.1 \\ \hline
\multicolumn{1}{l|}{\multirow{3}{*}{RL 2-cell}} & \multicolumn{1}{l|}{Power} &  & 1570 & 1981 & 1681 & 1879 & 1853 & 1884 & 2153 & 2495 & 1937 \\
\multicolumn{1}{l|}{} & \multicolumn{1}{l|}{QoS} &  & 100.0 & 100.0 & 100.0 & 100.0 & 100.0 & 99.7 & 99.7 & 92.2 & 98.9 \\
\multicolumn{1}{l|}{} & \multicolumn{1}{l|}{Handover} &  & 55.1 & 64.6 & 79.8 & 80.7 & 103.5 & 122.5 & 226.1 & 276.4 & 126.1 \\ \hline
\multicolumn{1}{l|}{\multirow{3}{*}{ADP 2-cell}} & \multicolumn{1}{l|}{Power} &  & 1567 & 1957 & 1678 & 1881 & 1845 & 1949 & 2080 & 2291 & 1906 \\
\multicolumn{1}{l|}{} & \multicolumn{1}{l|}{QoS} &  & 100 & 100 & 100 & 100 & 99.6 & 98.0 & 99.4 & 99.0 & 99.5 \\
\multicolumn{1}{l|}{} & \multicolumn{1}{l|}{Handover} &  & 25.6 & 69.3 & 111.1 & 79.8 & 152.9 & 283.1 & 356.2 & 518.7 & 199.6 \\ \hline
\addlinespace[-1.6pt]
 &  &  &  &  &  &  &  &  &  &  &  \\ \hline
\multicolumn{1}{l|}{\multirow{3}{*}{Rule-based 4-cell}} & \multicolumn{1}{l|}{Power} &  & 1194 & 1265 & 1419 & 1633 & 1639 & 1845 & 2160 & 2466 & 1702 \\
\multicolumn{1}{l|}{} & \multicolumn{1}{l|}{QoS} &  & 86.4 & 33.3 & 58.3 & 78.9 & 50.0 & 72.2 & 84.1 & 86.8 & 68.7 \\
\multicolumn{1}{l|}{} & \multicolumn{1}{l|}{Handover} &  & 95.9 & 265.0 & 560.5 & 263.1 & 947.1 & 633.6 & 496.8 & 419.9 & 460.2 \\ \hline
\multicolumn{1}{l|}{\multirow{3}{*}{ADP 4-cell}} & \multicolumn{1}{l|}{Power} &  & 1242 & 1622 & 1430 & 1508 & 1765 & 1729 & 1998 & 2396 & 1711 \\
\multicolumn{1}{l|}{} & \multicolumn{1}{l|}{QoS} &  & 100 & 98.9 & 97.9 & 100 & 98.0 & 98.61 & 98.2 & 92.56 & 98.0 \\
\multicolumn{1}{l|}{} & \multicolumn{1}{l|}{Handover} &  & 55.1 & 147.2 & 210.9 & 142.5 & 389.5 & 314.4 & 632.7 & 725.8 & 327.2 \\ \hline
\multicolumn{1}{l|}{\multirow{3}{*}{ADP Fixed 4-cell}} & \multicolumn{1}{l|}{Power} &  & 1540 & 1929 & 1615 & 1739 & 1847 & 1895 & 2056 & 2337 & 1869 \\
\multicolumn{1}{l|}{} & \multicolumn{1}{l|}{QoS} &  & 100.0 & 98.9 & 100.0 & 100.0 & 99.6 & 98.6 & 99.4 & 95.6 & 99.0 \\
\multicolumn{1}{l|}{} & \multicolumn{1}{l|}{Handover} &  & 38.9 & 74.1 & 153.9 & 159.6 & 171.0 & 169.1 & 276.4 & 917.7 & 245.1 \\ \hline
\end{tabular}
}
\end{table}

We test all the methods and present the results of the total power consumption, QoS, and total handovers in Table. \ref{tab:experiment_results} including the results per scenario and the average results across all scenarios. 
The compared baseline methods are identified as two types: one refers to 2-cell of which only the last two cells as switchable; the second refers to 4-cell, which permits the switching of up to four cells maintaining only the first cell active.
The rule-based and ADP methods can be set to in both 2-cell and 4-cell configurations. 
Figure. \ref{fig:methods_comparison}(a) depicts that the ADP 2-cell results in more power saved compared to the rule-based 2-cell method. Particularly, ADP 2-cell (1906 W) saves on average 11.0\% compared to \textit{No ES} (2141 W), while the rule-based 2-cell method (1967 W) reports 8.1\% energy saved.
Another rule-based method with the maximal switchable cells set to 4 cells reaches a similar power consumption of (1702 W) compared to ADP 4-cell (1711 W), 20.1\% power saved compared to \textit{No ES}. However, the QoS performance of the rule-based 4-cell is considerably lower than the other algorithms with an average QoS of 68.7\% shown in Figure \ref{fig:methods_comparison}(f) due to its over-ambitious switching depicted in Figure \ref{fig:methods_comparison}(e).
Rule-based methods are practicable when limited to a maximum of 2 cells switchable with a QoS of 98.5\%. However, the same rule that works well on the configuration of maximum 2 cells off applied to maximum 4 cells off will end up with the decisions causing a low QoS performance. The thresholds of rules are tunable to improve but lack the adaptability of the complexity of cell quantity.
In comparison, ADP-based methods demonstrate improved adaptability regarding different numbers of switchable cells.
The RL 2-cell shows a close albeit marginally lower performance of power consumption (1937 W) and a QoS (98.9\%) compared to ADP 2-cell (1906 W) and (99.5\%).  
ADP performs switching off cells following periodic traffic patterns and maintains as possible low number of active cells if no QoS compromise is observed as depicted in Figure. \ref{fig:methods_comparison}(b)(e).
As a result, ADP 4-cell obtains markedly improved power consumption with a marginally inferior, yet close QoS compared to the RL 2-cell or rule-based 2-cell.
Additionally, we provide one ablation test case: ADP Fixed with a fixed QoS threshold of 92. We conducted experiments of multiple fixed QoS thresholds in $[90,  95]$ and select 92 as being optimal in this range regarding balancing power consumption and QoS.
ADP Fixed obtained a lower average power consumption of $1869 W$ compared to ADP despite a marginally higher QoS. The ablation tests demonstrated the efficacy of the adaptive QoS threshold by online optimization.
Lastly, the number of handovers is highly correlated with the number of active cells switched to off, so \textit{No ES} keeps the lowest handovers.
Reducing power consumption will cause increased handovers. However, the trade-off arises when the switching power cost ($\Delta$ in Equation.  \ref{Eq:cost-to-go_functiohn}) is substantial. In our experiments, switching power cost was incorporated into the decision-making. which not necessarily reduce the handovers but make a contribution to the optimal power consumption.

\begin{figure}[!htbp]
    \centering
    \includegraphics[width=0.97\linewidth]{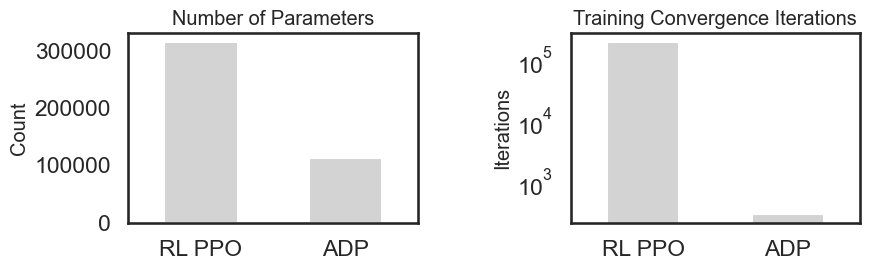}
    \caption{Comparison of the RL PPO and ADP models, referencing the results presented in Table \ref{tab:experiment_results}, illustrating the number of parameters and the training convergence iterations for each model with a batch size of 64.} 
    \label{fig_ppo_adp}
\end{figure}

As Figure. \ref{fig_ppo_adp} demonstrates, for the energy optimization trained on the same dataset of eight scenarios and produced the results in Table. \ref{tab:experiment_results}, ADP offers some benefits over PPO. 
The inherent design of ADP facilitates more straightforward training convergence, negating the extensive hyperparameters tuning typical of deep RL such as PPO. 
ADP also boasts a reduced computational cost with less number of parameters, ensuring memory efficiency. 
When ADP is provided with enough representative data encapsulating system dynamics, it can match the PPO's performance with more interpretable inference process, evident from the intermediate outcomes of its estimators.
\section{Conclusions}\label{conclusion}
Our work discussed the problem of reducing the energy consumption of wireless network deployments using ADP cell switching decisions.
Our proposed ADP methods tackle the trade-off between power consumption and maintaining the required QoS by combining neural network-based MLP and LSTM estimators for power consumption, QoS, and handover estimation to facilitate strategic action selection, thus allowing the method to adapt to changing real-time traffic conditions.
%
%
The ADP also exhibits adaptability in distinct dynamic traffic environments and ended up with a proper balance of power saved and the ratio of uncongested cells compared to various baseline methods.
The ADP scheme paired with neural network-based estimators can be broadly applicable to various wireless network configurations with different types of traffic scenarios and QoS metrics.
%


{
\bibliographystyle{IEEEtran}
\bibliography{ref.bib}
}
\end{document}